 \documentstyle{article}
\newfont{\g}{eufm9}

\newcommand{\gtg}{\mbox{\g g}}

\newcommand{\hgtg}{\mbox{$\hat{\gtg}$}}

\newcommand{\gtsl}{\mbox{\g sl}}

\newcommand{\gtn}{\mbox{\g n}}
\newcommand{\gtnp}{\gtn_{+}}
\newcommand{\gtnm}{\gtn_{-}}
\newcommand{\gth}{\mbox{\g h}}
\newcommand{\nc}{\mbox{${\bf C}$}}

\newcommand{\nz} {\mbox{${\bf Z}$}}
\newcommand{\nn} {\mbox{${\bf N}$}}

\newcommand{\cb}{\mbox{${\cal B}$}}
\newcommand{\cv}{\mbox{${\cal V}$}}

\newcommand{\cp}{\mbox{${\cal P}$}}

\newcommand{\cd}{\mbox{${\cal D}$}}

\newcommand{\cc}{\mbox{${\cal C}$}}
\newcommand{\ca}{\mbox{${\cal A}$}}

\newcommand{\bin}{\mbox{$\left( \begin{array}{c}n\\j \end{array}\right)$}}
\newcommand{\binq}{\mbox{$\left[ \begin{array}{c}n\\j \end{array}\right]$}}

\newtheorem{proposition}{Proposition}[section]

\newtheorem{theorem}{Theorem}[section]

\newtheorem{conjecture}[theorem]{Conjecture}

\newtheorem{lemma}[theorem]{Lemma}

\title{{ \bf  Rings of skew polynomials and Gel'fand-Kirillov conjecture for
quantum groups}}
\author{Kenji Iohara\thanks{e-mail address: iohara@kurims.kyoto-u.ac.jp},
Feodor
Malikov\thanks{Supported by the Japan Society for the Promotion of Science Post
-Doctoral Fellowship
for Foreign Researchers in Japan.} \thanks{Address after August, 1, 1993:
Mathematics Department Yale
University New Haven CT 06520 USA; e-mail address:
malikov@kusm.kyoto-u.ac.jp}\\
 Department of Mathematics, Kyoto University ,\\
 Kyoto 606 Japan }
\date{Received: }
 
\begin{document}

\maketitle
\pagestyle{myheadings}
\begin{abstract}
We introduce and study action of quantum groups on skew polynomial rings and
related
rings of quotients. This leads to a ``q-deformation'' of the Gel'fand-Kirillov
conjecture which
we
partially prove. We propose a construction of automorphisms of certain
non-commutaive rings of
quotients coming from complex powers of quantum group generators; this is
applied to explicit
calculation of singular vectors in Verma modules over $U_{q}(\gtsl_{n+1})$.
 We finally give a definition of a $q-$connection with coefficients in a ring
of
skew polynomials and study the structure of quantum group modules twisted by a
$q-$connection.
\vspace{5 mm}

\end{abstract}
\section {{\bf Introduction}}
\markright{ Skew polynomials and quantum groups}

This work was mainly inspired by the Feigin's construction which associates to
an element of the
Weyl group $w\in W$ an associative algebra homomorphism of a ``nilpotent part''
of a quantum group
to an appropriate algebra of skew polynomials:
\begin{equation}
\label{intr_feig_morph}
\Phi(w):\;U_{q}^{-}(\gtg)\rightarrow \nc[X],
\end{equation}
where $X$ stands for $X_{1},\ldots,X_{l}$, $l$ is a length of $w$ and
$X_{j}X_{i}=q^{\alpha_{ij}}X_{i}X_{j}$ for some $\alpha_{ij}\in\nz,\;1\leq
i,j\leq l$.
The main topics treated in the work are as follows.

{\bf 1.}{\em Realizations of Lie algebras and quantum groups and
Gel'fand-Kirillov conjecture.}
The fact that a  Lie algebra of an algebraic group (``algebraic Lie algebra'')
 can be realized in differential
operators acting on a suitable manifold is, probably, more fundamental than the
notion of a Lie
algebra. Explicit formulas for such a realization in the case when the algebra
is simple, the manifold
is a big cell of a flag manifold have become especially popular recently
because of their relation
to the free field approach to 2-dimesional conformal field theory (
{}~\cite{feig_fren_0,bowkn1}).
An important property
of the realization was discovered by Gel'fand and Kirillov ~\cite{gk1} long ago
and in a remarkable generality. Their observation is that however
complicated classification of algebraic Lie algebras may be, equivalence
classes of rings of
quotients  of universal enveloping algebras are labelled by pairs of positive
integers:
a ring of
quotients of a universal enveloping algebra is isomorphic with a ring of
quotients of a ring of
differential operators on $n$ variables with polynomial coefficients
trivially extended by a $k-$dimensional center, where  $k$ is a
dimension of a generic orbit in the coadjoint representation and
$2n+k$ is equal to the dimension of the algebra . (This conjecture has been
proven by
themselves and others  in many cases ~\cite{gk1,jos,mcc}.)

A natural class of rings suitable for formation of  rings of quotients is
provided by the so-called
{\em Ore domains}. Besides above mentioned universal enveloping algebras of
finite dimensional Lie
algebras and rings of differential operators the class of Ore domains comprises
(deformed)
enveloping algebras of affine Lie algebras, rings of skew polynomials
and $q-$difference operators. We prove that the Feigin's
morphism $\Phi(w_{0})$ associated to the longest Weyl group element provides an
isomorphism
of rings of quotients $Q(U_{q}^{-}(\gtsl_{n+1}))\approx Q(\nc[X])$. This
isomorphism allows to equip
$Q(\nc[X])$ with a structure of $U_{q}(\gtsl_{n+1})$-module. More precisely we
define an
$n-$parameter family of associative algebra homomorphisms from
$U_{q}(\gtsl_{n+1})$ to an algebra of
$q-$difference operators with coefficients in $Q(\nc[X])$. We conjecture that
this provides an
isomorphism of $Q(U_{q}(\gtsl_{n+1}))$ with an $n-$dimensional central
extension of the algebra
of $q-$difference operators. We prove this conjecture for
$U_{q}(\gtsl_{2}),\;U_{q}(\gtsl_{3})$
in a slightly weaker form.

{\bf 2.} {\em Complex powers, automorphisms and screening operators.} A
remarkable observation made
in early works on Kac-Moody algebras ~\cite{lepwil,fk} is
that affine Lie algebras, like finite dimensional
simple ones, are also realized in differential operators, though  on infinite
many variables. A
 family of such realizations depending on a highest weight $\lambda$ was
constructed by Wakimoto
{}~\cite{wak}
for $\widehat{\gtsl_{2}}$ and by Feigin and Frenkel ~\cite{feig_fren_0}
for all non-twisted affine Lie algebras.
Thus obtained modules are now known as {\em Wakimoto modules} $F(\lambda)$. The
main ingredient of the
2-dimensional conformal field theory associated to an affine algebra $\hgtg$ is
a 2-sided complex
consisting of direct sums of Wakimoto modules
\[\cdots\rightarrow F^{(-1)}\rightarrow F^{(0)}\rightarrow
F^{(1)}\rightarrow\cdots\]
such that its homology is concentrated in the 0-th dimension and is equal to an
irreducible highest
weight module over $\hgtg$ ( {\em BRST resolution}).

Bowknegt, McCarthy and Pilch revealed a quantum group structure hidden in the
differential of the
BRST  resolution. Recall that $U_{q}(\gtg)$-morphisms of a Verma module
$M(\lambda)$ into a Verma
module $M(\mu)$ are in 1-1 correspondence with {\em singular vectors } of the
weight $\lambda$ in
the latter ( the correspondence is established by assigning to a morphism an
image of the vacuum
vector under this morphism). Denote by $Sing_{\lambda}(M(\mu))$ the set of
singular vectors of the
weight $\lambda$ in $M(\mu)$. It is argued in ~\cite{bowkn1,bowkn2} that there
is a linear map
\begin{equation}
\label{intr_b_mc_p}
Sing_{\lambda}(M(\mu))\rightarrow Hom_{\hgtg}(F(\lambda),\,F(\mu))
\end{equation}
and that conjecturally this map is an isomorphism.

Singular vectors in Verma modules over quantum groups related to an arbitrary
Kac-Moody algebra
were found in the form ~\cite{malff,mal}
\begin{equation}
\label{intr_s_vect_gener}
F_{i_{l}}^{s_{l}}\cdots F_{i_{1}}^{s_{1}}F_{i_{0}}^{N}F_{i_{1}}^{t_{1}}\cdots
F_{i_{l}}^{t_{l}},
\end{equation}
where $s_{i},t_{i},\;1\leq i\leq l$ are appropriate complex numbers,
$F_{i},\;1\leq i\leq n$ are
canonical Cartan generators of $U_{q}^{-}(\gtg)$, $N\in\nn$. Here we carry out
an explicit calculation
of (~\ref{intr_s_vect_gener}), i.e. rewrite it in the form containing only
natural powers of $F'$s.
Observe that the map (~\ref{intr_b_mc_p}) is determined by
assigning to each  $F_{i}$  what is known as a
{\em screening operator}.

One of the consequences of the prescription how to choose powers in
(~\ref{intr_s_vect_gener}) is that
$s_{i}+t_{i},\;1\leq i\leq l$ are all non-negative integers. More generally,
one may consider a map
\[U_{q}^{-}(\gtg)\ni p\mapsto F_{i}^{\beta}pF_{i}^{-\beta},\;\beta\in\nc.\]
A simple calculation using the notion of the $q-${\em commutator} shows that
this map extends to an
automorphism of the quotient ring $Q(U_{q}^{-}(\gtg))$. Therefore a singular
vector is obtained by,
roughly speaking,
 a sequence of automorphisms applied to a Cartan generator.

Similarly one may consider an operator of conjugation by a complex power of a
linear form
\[\nc[x_{1},\ldots,x_{k}]\ni p\mapsto (x_{i_{1}}+\cdots x_{i_{l}}
)^{\beta}p(x_{i_{1}}+\cdots x_{i_{l}}
)^{-\beta},\]
acting on a certain completion of a ring of skew polynomials
$\nc[x_{1},\ldots,x_{k}]$. Simple but
nice calculation based on the $q-$binomial theorem shows that this map actually
determines an
automorphism of the ring of quotients $Q(\nc[x_{1},\ldots,x_{k}])$. This
construction may be
interesting in its own right: unlike the things are in commutative realm, the
very existence of
( non-trivial ) automorphisms of $Q(\nc[x_{1},\ldots,x_{k}])$ is not quite
obvious. Combined with the
Feigin's morphism this construction answers an informal question: ``how does it
happen that complex
powers in the singular vector formula cancel out?'' Another application of
these automorphisms is
that they produce natural examples of $q-$connections with coefficients in skew
polynomials.

{\bf 3.} {\em Quantum group modules twisted by $q-$connections.} It has been
realized
{}~\cite{awata,f_g_p_p} that
the ``singular vector decoupling condition'' makes it necessary to consider
non-bounded --
neither highest nor lowest weight -- modules in 2-dimensional conformal field
theory at a rational
level. On the other hand the singular vector formula (~\ref{intr_s_vect_gener})
makes it natural to
consider an extension of a Verma module by complex powers of generators, which
transparently
produces non-bounded modules. It was shown in ~\cite{feig_mal} that the duals
to such modules
 are realized in
multi-valued functions on a flag manifold or, in other words, in modules
twisted by connections; in
particular a family of integral intertwining operators acting among such
modules was constructed.
( One may also find in ~\cite{feig_mal}
 and in the forthcoming paper ~\cite{iohm} integral formulas for solutions to
Knizhnik-Zamolodchikov equations with coefficients in non-bounded modules.)

Here we adjust the definition of a $q-$connection given by Aomoto and Kato
{}~\cite{aomok} in the commutative case
to the case of skew polynomials. This definition identifies $q-$connections
with the cohomology
group $H^{1}(\nz^{k},\;Q(\nc[x_{1},\ldots,x_{k}]))$. We also produce a family
of elements of
$H^{1}(\nz^{k},\;Q(\nc[x_{1},\ldots,x_{k}]))$ associated with complex powers of
linear forms all
this being independent of quantum groups. In the case when the ring of skew
polynomials is the one
coming from the Feigin's morphism, the twisting by such a $q-$connection is
nothing but
a passage from a Verma module to its extension by complex powers of generators.
This allows to
construct $q-$analogues of the intertwiners of ~\cite{feig_mal},
which in the quantum case may be thought of as
right multiplications by certain complex powers of linear forms. We also find
out what the
$U_{q}(\gtg)-$module structure of a Verma module extended by a complex power of
only 1 generator is.
Such a module can be viewed as a module induced from a non-bounded module over
a parabolic
subalgebra. It turns out that its structure is formally close to that of a
Verma module. In
particular, the notion of a singular vector is naturally replaced with that of
a {\em singular chain}
and a singular chain encodes information on a family of singular vectors.

{\bf Acknowledgments.} Our thanks go to B.Feigin who explained to us his
unpublished results,
to N.Reshetikhin who brought to our attention the paper ~\cite{aomok} and to
M.Jimbo for his interest in the work. Some results of the work were reported at
Mie University in Tsu.
 F.M. is indebted to
M.Wakimoto for his heartiest hospitality during the visit.

\section{{\bf Main definitions related to quantum groups}}
 The material of this secton is fairly standard. Usually the reference is the
work ~\cite{dc_kac}.

{\bf 1.} Let,  as usual, $A=(a_{ij}),\,1\leq i,j\leq n$ stand for a generalized
symmetrizable
Cartan matrix ,
symmetrized by non-zero relatively prime integers $d_{1},\ldots, d_{n}$
 such that $d_{i}a_{ij}=d_{j}a_{ji}$ for all $i,j$. A Kac-Moody Lie algebra
$\gtg$
attached to $A$ is an algebra on
generators $E_{i}, F_{i}, H_{i}, \,1\leq i\leq n$ and well-known relations
explicitly depending on
 entries of $A$ ( see ~\cite{kac_book} ). Among the structures related to
$\gtg$ we shall use the
following:

the triangular decomposition $\gtg=\gtnm\oplus\gth\oplus\gtnp$;

the dual space $\gth^{\ast}$; elements of  $\gth^{\ast}$ will be referred to as
weights;

the root space decomposition
$\gtn_{\pm}=\oplus_{\alpha\in\Delta_{\pm}}\gtg_{\alpha}$,
$\gtg_{\alpha_{i}}=\nc E_{i}$;

the root lattice $Q\in\gth^{\ast}$,
$\{\alpha_{1},\ldots,\alpha_{n}\}\subset\Delta_{+}\subset\gth^{\ast}$ being the
set of simple roots;

the invariant bilinear form $Q\times Q\rightarrow\nz$ defined by
$(\alpha_{i},\alpha_{j})=d_{i}a_{ij}$.

 {\bf 2.}For $q\in\nc,\;d\in\nz$ set:
\[[n]_{d}=\frac{1-q^{2nd}}{1-q^{2d}},\]
\[[n]_{d}!=[n]_{d}\cdots [1]_{d},\]
\[\binq_{d} = \frac{[n]_{d}\cdots [n-j+1]_{d}}{[j]_{d}!},\]
 omitting the subscript if $d=1$.

Suppose $\gtg$ is a Kac-Moody Lie algebra  attached to $A$.
{\em The Drinfeld -Jimbo quantum group $U_{q}(\gtg),\;q\in\nc$}
is said to be
a Hopf algebra with antipode $S$, comultiplication $\Delta$
 and 1 on
generators $E_{i}, F_{i}, K_{i}, K_{i}^{-1},\,1\leq i\leq n$ and defining
relations

\begin{equation}
\label{q_1}
K_{i}K_{i}^{-1}=K_{i}^{-1}K_{i}=1,\;\;K_{i}K_{j}=K_{j}K_{i},
\end{equation}
\begin{equation}
\label{q_2}
%% FOLLOWING LINE CANNOT BE BROKEN BEFORE 80 CHAR
%% FOLLOWING LINE CANNOT BE BROKEN BEFORE 80 CHAR
K_{i}E_{j}K_{i}^{-1}=q^{a_{ij}}_{i}E_{j},\;\;K_{i}F_{j}K_{i}^{-1}=q^{-a_{ij}}_{i}F_{j},\;
q_{i}=q^{d_{i}},
\end{equation}
\begin{equation}
\label{q_3}
E_{i}F_{j}-F_{j}E_{i}=\delta_{ij}\frac{K_{i}-K_{i}^{-1}}{q_{i}-q_{i}}
,\; q_{i}=q^{d_{i}},
\end{equation}

\begin{eqnarray}
\sum_{\nu =0}^{1-a_{ij}}(-1)^{\nu}q_{i}^{\nu(\nu - 1+ a_{ij})}
\left[ \begin{array}{c}1-a_{ij}\\\nu \end{array}\right]_{d_{i}}
E_{i}^{1-a_{ij}-\nu}E_{j}E_{i}^{\nu} & = & 0,\;\; \nonumber\\
\sum_{\nu =0}^{1-a_{ij}}(-1)^{\nu}q_{i}^{\nu(\nu - 1+ a_{ij})}
\left[ \begin{array}{c}1-a_{ij}\\\nu \end{array}\right]_{d_{i}}
F_{i}^{1-a_{ij}-\nu}F_{j}F_{i}^{\nu} & = & 0\;\;(i\neq j) \label{q_4}
,
\end{eqnarray}

the comultiplication being given by
\begin{equation}
\label{comult}
\Delta E_{i}=E_{i}\otimes 1+K_{i}\otimes E_{i},
\Delta F_{i}=F_{i}\otimes K_{i}^{-1}+1\otimes F_{i},
\Delta K_{i}=K_{i}\otimes K_{i},
\end{equation}
and antipode - by
\begin{equation}
\label{antip}
S E_{i}=-K_{i}^{-1}E_{i},
S F_{i}=-F_{i}K_{i},
S K_{i}=K_{i}^{-1}.
\end{equation}

The relations admit the $\nc-$algebra anti-automorphism $\omega$
\begin{equation}
\label{autom_q}
\omega E_{i}=F_{i},\;\omega F_{i}=E_{i},\;\omega K_{i}=K_{i}
\end{equation}

 Set $U^{+}_{q}(\gtg)\;
(U^{-}_{q}(\gtg))$
 equal to the subalgebra, generated by $E_{i}\;(F_{i}\; resp.)\;(1\leq i\leq
n)$
and $U^{0}_{q}(\gtg)=\nc[K_{1}^{\pm 1},\ldots
,K_{n}^{\pm 1}]$. We will sometimes reduce these notations to $U_{q}^{+},
U_{q}^{-}, U_{q}^{0}$
if this does leads to confusion.
One may check  that the multiplication induces an isomorphism of linear spaces
\begin{equation}
\label{triang_decomp_q}
U_{q}(\gtg)\approx U^{-}_{q}(\gtg)\otimes U^{0}_{q}(\gtg)\otimes
U^{+}_{q}(\gtg).
\end{equation}

Set $U_{q}^{\geq}=U_{q}^{0}U_{q}^{+} $.
{}From now on unless otherwise stated $A$ is assumed to be of finite type.

{\bf 3}. For any $Q-$ graded associative algebra $\ca = \oplus_{\beta\in Q}
\ca_{\beta}$ define a
 {\em $q-$commutator}, which associates to any homogeneous $b\in\ca_{\beta}$ a
mapping
\[ ad_{q}b\,:\;\ca\rightarrow \ca\]
of degree $\beta$ determined by
\begin{equation}
\label{q_bracket}
ad_{q}b(c)=bc-q^{(\beta,\gamma)}cb,\;\mbox {if}\; c\in\ca_{\gamma}.
\end{equation}
One deduces that the $q-$bracket is a $q-$derivation of $\ca$, meaning that
\begin{equation}
\label{q_derivation}
ad_{q}a(bc)=(ad_{q}a(b))c+q^{(\alpha,\beta)}b\,ad_{q}a(c)\;
\mbox{if}\;a\in\ca_{\alpha},b\in\ca_{\beta},c\in\ca_{\gamma}.
\end{equation}

Using (\ref{q_derivation}) one proves the following useful formula
\begin{equation}
\label {move_right}
b^{n}c=q^{n(\beta,\gamma)} cb^{n}+\sum_{j=1}^{n} q^{(n-j)(\beta,\gamma)}
\binq_{(\beta,\beta)/2} (ad_{q}b)^{j}(c)b^{n-j},
\end{equation}
its Lie algebra analogue being

\begin{equation}
\label {move_right_class}
b^{n}c= cb^{n}+\sum_{j=1}^{n}
\bin (ad\,b)^{j}(c)b^{n-j}.
\end{equation}

In the case $\ca=U_{q}^{\pm}(\gtg)$ one realizes that the relations (\ref{q_4})
simply mean that
\begin{equation}
\label{q_4_mod}
(ad_{q}E_{i})^{-a_{ij}+1}(E_{j})=(ad_{q}F_{i})^{-a_{ij}+1}(F_{j})=0,\mbox{ if
}i\neq j.
\end{equation}

The following observation will be used below in the discussion of rings of
quotients:

the relations (\ref{q_derivation},\ref{move_right},\ref{q_4_mod}) imply that
for any $F_{i},
b\in U_{q}^{-}(\gtg)$ one has
\begin{equation}
\label{left_mult_simpl}
F_{i}^{N}b\in U_{q}^{-}(\gtg)F_{i},
\end{equation}
for all sufficiently large $N$.

{\bf 4.} Following Lusztig ~\cite{lus} introduce the following automorphisms
$R_{i},\;1\leq i\leq n$
 of the algebra $U(\gtg)$:
\begin{eqnarray}
R_{i}E_{i}=-F_{i}K_{i},\;
R_{i}E_{j}&=&\sum_{s=0}^{-a_{ij}}(-1)^{s-a_{ij}}q_{i}^{-s}
%% FOLLOWING LINE CANNOT BE BROKEN BEFORE 80 CHAR
%% FOLLOWING LINE CANNOT BE BROKEN BEFORE 80 CHAR
\frac{E_{i}^{-a_{ij}-s}}{[-a_{ij}-s]_{d_{i}}!}E_{j}\frac{E_{i}^{s}}{[s]_{d_{i}}!}
\label {luszt_aut_1}\\
\mbox{ if } i&\neq& j,\nonumber\\
R_{i}F_{i}=-K_{i}^{-1}E_{i},\;
R_{i}F_{j}&=&\sum_{s=0}^{-a_{ij}}(-1)^{s-a_{ij}}q_{i}^{s}
%% FOLLOWING LINE CANNOT BE BROKEN BEFORE 80 CHAR
%% FOLLOWING LINE CANNOT BE BROKEN BEFORE 80 CHAR
\frac{F_{i}^{s}}{[s]_{d_{i}}!}E_{j}\frac{E_{i}^{-a_{ij}-s}}{[-a_{ij}-s]_{d_{i}}!}
\label {luszt_aut_2}\\
\mbox{ if } i&\neq j&,\nonumber\\
R_{i}K_{j}&=&K_{j}K_{i}^{-a_{ij}}\label {luszt_aut_3}.
\end{eqnarray}

Fix a reduced decomposition $r_{i_{1}}r_{i_{2}}\cdots r_{i_{N}}$ of the longest
element of the
Weyl group $W$. This gives an ordering of the set of positive roots:
\[\beta_{1}=\alpha_{i_{1}},\;\beta_{2}=r_{i_{1}}\alpha_{i_{2}},\ldots ,
\beta_{N}=r_{i_{1}}
\cdots r_{i_{N-1}}\alpha_{i_{N}}.\]
One introduces root vectors ~\cite{lus}
\begin{eqnarray}
 E_{\beta_{s}}=R_{i_{1}}\cdots R_{i_{s-1}}E_{i_{s}}\label{def_root_vect_1},\\
F_{\beta_{s}}=R_{i_{1}}\cdots R_{i_{s-1}}F_{i_{s}}\label{def_root_vect_2}.
\end{eqnarray}
For $k=(k_{1},\ldots , k_{N})\in \nz^{N}_{+}$ set
$E^{k}=E^{k_{1}}_{\beta_{1}}\cdots E^{k_{N}}_{\beta_{N}},\;F^{k}=\omega E^{k}.$

 \begin{proposition}

(i) ~\cite{lus} Elements $E^{k}\,(F^{k}\mbox{ resp. }),\, k\in\nz^{N}_{+}$,
form a basis
 of $U_{q}^{+}(\gtg)$
($U_{q}^{-}(\gtg)$ resp.) over $\nc$.

(ii) ~\cite{dc_kac} The algebra $U_{q}(\gtg)$ affords a structure of a
$\nz^{2N+1}_{+}-$filtered algebra, so that the associated graded algebra
$Gr(U_{q}(\gtg))$ is
an associative algebra over $\nc$ on generators $E_{\alpha},\,
F_{\alpha}\;(\alpha\in\Delta^{+})\;
K_{i}^{\pm}\; (0\leq i \leq n)\;$ subject to the following relations
\[K_{i}K_{j}=K_{j}K_{i},\;K_{i}K_{i}^{-1}=1,\;
E_{\alpha}F_{\beta}=F_{\beta}E_{\alpha},\]
\[K_{i}E_{\alpha}=q^{(\alpha,\alpha_{i})}E_{\alpha}K_{i},
K_{i}F_{\alpha}=q^{-(\alpha,\alpha_{i})}F_{\alpha}K_{i},\]
\[E_{\alpha}E_{\beta}=q^{(\alpha,\beta)}E_{\beta}E_{\alpha},
F_{\alpha}F_{\beta}=q^{(\alpha,\beta)}F_{\beta}F_{\alpha},\mbox{ if } \alpha >
\beta.\]
\label{filter_prop}
\end{proposition}
Recall that an algebra $\nc^{s}[x_{1},\ldots ,x_{k}]$ on generators
$x_{1},\ldots ,x_{k}$ and
defining relations $x_{i}x_{j}=\lambda_{ij}x_{j}x_{i}$ for $i>j$, where
$\lambda_{ij}\in\nc^{\ast}$,
is called {\em an algebra of skew polynomials}. Therefore, the item (ii) of
Proposition
\ref{filter_prop} asserts that $Gr(U_{q}(\gtg))$ is a skew polynomial algebra.
The ``classical''
analogue of this is the fact that $Gr(U(\gtg))$ is a symmetric algebra
$S(\gtg)$.

An  algebra of skew polynomials has no zero divisors, therefore, the same is
true for $U_{q}(\gtg)$
{}~\cite{dc_kac}.

\section { {\bf Rings of quotients associated to (deformed) enveloping
algebras}}

\subsection { {\bf Gelgand - Kirillov' conjecture and Feigin's construction.}}
\label{gel_kir_feig}

{\bf 1.} A ( non-commutative )
ring $\ca$ with no zero divisors
is called an {\em Ore domain} if  any 2 elements of $\ca$ have a common right
and a common
left multiple. A class of examples of Ore domains is provided by the rings of
polynomial growth.
We shall be calling an $\nn-$ filtered ring $\ca=\cup_{i\geq 1}\ca^{(i)}$ a
ring of polynomial
growth if $dim \ca^{(i)}$ is equivalent to a certain polynomial as
$i\rightarrow \infty$.

\begin{lemma}
\label{ore-dom}
A ring of polynomial growth with no zero divisors is an Ore domain.
\end{lemma}

{\em Proof.} Assume that $dim \ca^{(i)}\sim a_{0}i^{k},$ as $i\rightarrow
\infty$. Then for any ideal
$I$ on 1 generator one has: $dim I^{(i)}\sim
a_{0}i^{k},\;I^{(i)}=I\cap\ca^{(i)}$. If 2 ideals
$I_{1},I_{2}$ on 1 generator have zero intersection then $
(I_{1}+I_{2})^{(i)}\sim 2a_{0}i^{k}$, contradicting the assumption. $\Box$

The simplest examples of rings of
polynomial growth are, therefore,  algebras of (skew) polynomials. Further,
affine and
finite-dimensional Lie algebras are distinguished among Kac-Moody algebras as
algebras of polynomial
growth ~\cite{kac_book}.
 This combined with Proposition \ref{filter_prop} implies that $U(\gtg)$, if
$\gtg$ is of
either affine or finite dimensional, and $U_{q}(\gtg)$, if $\gtg$ is finite
dimensional,  are Ore
domains, as well as the corresponding $ U^{\pm}(\gtg)\;(U^{\pm}_{q}(\gtg)$.

{\bf 2.} An Ore domain $\ca$ is a suitable object for formation of a ring of
quotients. Consider
expressions of the form $ab^{-1},\;b^{-1}a,\;a,b\in\ca$ called right and left (
resp.) quotients.
Introduce a relation $\approx$ by saying that
\[ \mbox{ (i) }ab^{-1}\approx c^{-1}d\Leftrightarrow ca=db,\]
(ii) 2 right (left) quotients are in relation $\approx$ if and only if they are
in relation
$\approx$ to one and the same left (right) quotient.

The Ore domain conditions imply that $\approx$ is an equivalence relation.
Denote
the set of equivalence classes of $\approx$ by $Q(\ca)$. One more application
of the Ore domain
conditions gives that each equivalence class contains left and right quotients
and that any 2
left ( right ) quotients are eqivalent to left (right) quotients with one and
the same denominator.
This allows to define operations of addition and multiplication ( in the most
natural way ), which
completes the definition of the ring of quotients $Q(\ca)$.

A definition of a ring of quotients $\ca [S^{-1}]$ with respect to $S\subset
\ca$ is a more subtle
matter because due to the noncommutativity of $\ca$ it is not clear what can
really appear as a
denominator. However in the case when $\ca$ is a (quantized) enveloping algebra
one can say more.
It follows from
(\ref{move_right}) that
\[ F_{i}^{-n}F_{j}=q^{-n(\alpha_{i},\alpha_{j})}
F_{j}F_{i}^{-n}+\sum_{j=1}^{\infty}
q^{(-n-j)(\alpha_{i},\alpha_{j})} \left[ \begin{array}{c}-n\\j
\end{array}\right]_{(\alpha_{i},\alpha_{i})/2} (ad_{q}F_{i})^{j}(F_{j})F_{i}
^{-n-j},\]
if $ i\neq j,\;n>0.$

(Observe that (\ref{q_4_mod}) implies that only finite number of terms in the
right-hand
side of the above formula can be non-zero and, therefore, it makes sense as an
element of
$U_{q}^{-}(\gtg)$.) Therefore, the result of commuting negative powers of a
Cartan generator to the
right is negative powers of the same generator on the right. One also deduces
from
(\ref{left_mult_simpl}) that 2 words
$F_{i_{1}}^{s_{1}}\cdots F_{i_{m}}^{s_{m}}$, $F_{j_{1}}^{t_{1}}\cdots
F_{i_{l}}^{t_{l}}$ have
a common right multiple of the form
\[F_{i_{1}}^{N_{1}}\cdots F_{i_{m}}^{N_{m}}F_{j_{1}}^{t_{1}}\cdots
F_{i_{l}}^{t_{l}},\]
 if
$N_{1},\ldots , N_{m}$ are sufficiently large.
Now, if $S\subset U_{q}^{-}(\gtg)$
is a multiplicatively
closed subset multiplicatively generated by $F_{i_{1}},\ldots ,F_{i_{k}}$,
 one defines $U_{q}^{-}(\gtg)[S^{-1}]$ as a subset of $Q(U_{q}^{-}(\gtg))$
consisting of classes
of quotients of the form $ab^{-1},\;b\in S$. The above discussion shows that
$U_{q}^{-}(\gtg)[S^{-1}]$
is a subring. We shall sometimes denote $U_{q}^{-}(\gtg)[S^{-1}]$ by
$U_{q}^{-}(\gtg)[F_{i_{1}}^{-1}, F_{i_{2}}^{-1},\ldots ,F_{i_{k}}^{-1}]$.

It is easy to see that the same goes through with $F'$s replaced with $E'$s or
$U_{q}^{-}(\gtg)$
replaced with $U_{q}(\gtg)$, as well as with everything replaced with its
classical ($q\rightarrow
1$) analogues. Further, even though a ring of quotients is not defined for an
arbitrary Kac-Moody
algebra, this discussion shows that a ring of quotients $U(\gtg)[S^{-1}]$ is
well-defined if
$S$ is multiplicatively generated by a (sub)set of real root vectors. Actually,
formulas
(\ref{q_derivation},\ref{move_right_class} ) provide an algorithm of carrying
out operations of
multiplication and addition on elements of $U(\gtg)[S^{-1}]$.

{\bf 3.} It often happens that  rings of quotients of universal enveloping
algebras
of different finite-dimensional Lie algebras are isomorphic with each other.
Denote by
$D_{n k}$ an algebra on generators $a_{1},\ldots ,a_{n},a_{1}^{\ast},\ldots
,a_{n}
^{\ast},c_{1},\ldots, c_{k}$ and defining relations
\[
[a_{i},a_{j}^{\ast}]=\delta_{ij},\;[c_{i},a_{j}]=[c_{i},a_{j}^{\ast}]=[a_{i},
a_{j}]=[a_{i}^{\ast},a_{j}^{\ast}]=0,\;\mbox{for
all }i,j.\] $D_{nk}$ can, of course, be viewed as an algebra of differential
operators on $n$
variables trivially extended by $k-$dimensional center.

\begin{conjecture} (\mbox{ Gelfand - Kirillov ~\cite{gk1} }) If $\gtg$ is an
algebraic Lie algebra then
$Q(U(\gtg))$ is isomorphic with $Q(D_{nk})$ for $k$ equal to the dimension of a
generic $\gtg-$orbit
in the coadjoint representation and $n=(dim\gtg - k)/2$. \end{conjecture}

This conjecture has been proven in many cases ~\cite{gk1,jos,mcc}.

{\bf 4.} It seems that the following construction ( due to Feigin ~\cite{feig}
)
 is relevant to a proper
$q-$deformation of the Gelfand-Kirillov's conjecture. For a pair of $Q-$graded
associative algebras
$\ca,\;\cb$
define a {\em $q-$twisted tensor product} as an algebra $\ca\otimes_{q}\cb$
isomorphic with
$\ca\otimes \cb$ as a linear space and with the multiplication given by
$a_{1}\otimes b_{1}\cdot
 a_{2}\otimes b_{2}=q^{(\alpha , \beta)}a_{1}a_{2}\otimes b_{1}b_{2}$ if
$a_{2}\in\ca^{(\alpha)},\;
b_{1}\in\cb^{(\beta)}$. Evidently, $\ca\otimes_{q}\cb$ is a $Q-$graded algebra.

\begin {proposition} ~\cite{feig} For any Kac-Moody algebra $\gtg$ the map
\[ \tilde{\Delta}:\;U_{q}^{\pm}(\gtg)\rightarrow U_{q}^{\pm}(\gtg)\otimes_{q}
U_{q}^{\pm}(\gtg),\]
\[\tilde{\Delta}:1\mapsto 1\otimes 1,\]
\[\tilde{\Delta}:
E_{i}\mapsto E_{i}\otimes 1 +1\otimes E_{i}\;(F_{i}\rightarrow F_{i}\otimes 1
+1\otimes F_{i}
\mbox{ resp.}),\]
$1\leq i\leq n$

is a homomorphism of associative algebras.
\label{feig_comult}
\end{proposition}

{\em Remark.} It is known that the map
$U_{q}^{\pm}(\gtg)\rightarrow U_{q}^{\pm}(\gtg)\otimes U_{q}^{\pm}(\gtg)$  does
not exist in the
category of associative algebras.

Iterating $\tilde{\Delta}$ one obtains a sequence of maps
\[\tilde{\Delta}^{m}:U_{q}^{-}(\gtg)\rightarrow U_{q}^{-}(\gtg)^{\otimes
m},\;m=2,3\ldots,\]
determined by $\tilde{\Delta}^{2}=\tilde{\Delta},
 \tilde{\Delta}^{m}=(\tilde{\Delta}\otimes id)\circ \tilde{\Delta}^{m-1}$.

For any simple root $\alpha_{i}$ consider a ring of polynomials on 1 variable
$\nc [X_{i}]$,
which we regard as
 a $Q-$graded by setting $deg\,X_{i}=\alpha_{i}$. There arises a morphism of
$Q-$graded associative
algebras
\[\rho_{i}:\;U_{q}^{-}(\gtg)\rightarrow \nc[X_{i}],\]
\[F_{j}\mapsto \delta_{ij} x_{i}.\]

Now, for any sequence of simple roots $\alpha_{i_{1}},\ldots ,\alpha_{i_{k}}$
there arises a
morphism of $Q-$graded associative algebras:
\[(\rho_{i_{1}}\otimes\cdots\otimes \rho_{i_{k}})\circ\tilde{\Delta}^{k}:\;
U_{q}^{-}(\gtg)\rightarrow
\nc[X_{1i_{1}}]\otimes_{q}\cdots\otimes_{q}\nc[X_{ki_{k}}].\]
 ( The double indexation of $X'$s
is necessary because some number can appear in the sequence $i_{1},\ldots
,i_{k}$ more than once but
the corresponding indeterminates have to be regarded as different.)

Evidently, $\nc[X_{1i_{1}}]\otimes_{q}\cdots\otimes_{q}\nc[X_{ki_{k}}]$
is an algebra of
skew polynomials $\nc[X_{1i_{1}}\ldots X_{ki_{k}}]$,
 satisfying the relations
$X_{si_{s}}X_{ti_{t}}=q^{(\alpha_{i_{s}},\alpha_{i_{t}})}
X_{ti_{t}}X_{si_{s}},\;s>t$. Therefore, we have constructed a family of
morphisms of
a `` maximal nilpotent subalgebra '' of a quantum group associated to an
arbitrary Kac-Moody algebra
to algebras of skew polynomials.
It is interesting that a proper classical analogue of this construction is not
so obvious and is
best understood in the framework of rings of quotients ( see below).

We now assume that $\gtg$ is a simple finite-dimensional Lie algebra. Let
$w_{0}=r_{i_{1}}\cdots
r_{i_{N}}\in W$ be a reduced decomposition of the element of maximal
 length. Set $\Phi(i_{1},\ldots ,i_{N})=
(\rho_{i_{1}}\otimes\cdots\otimes \rho_{i_{N}})\circ\tilde{\Delta}^{N}$.

\begin{conjecture} ~\cite{feig}
\label {conj_of_feig}

(i)  $\Phi(i_{1},\ldots ,i_{N})$ is an embedding.

(ii) $\Phi(i_{1},\ldots ,i_{N})$ extends -- at least for a special choice of a
reduced decomposition
$w_{0}=r_{i_{1}}\cdots
r_{i_{N}}$ --
to an isomorhism of $Q(U_{q}^{-}(\gtg))$ with
$Q(\nc[X_{1i_{1}}\ldots X_{Ni_{N}}])$.
\end{conjecture}

{\bf 5.} {\em Example: $\gtg=\gtsl_{n+1}$.} From the abstract point of view
$\gtg=\gtsl_{n+1}$ is an
algebra related to the Cartan matrix $(a_{ij})$, where
\[ a_{ij}= \left\{ \begin{array}{rll} 2 &\mbox{if}&i=j\\
                                    -1 &\mbox{if}& \mid i-j\mid =1\\
                                    0 &\mbox{if}& \mid i-j\mid >1
\end{array}\right.\]

Choose a reduced decomposition of the longest Weyl group element to be
$w_{0}=r_{1}r_{2}\cdots
r_{n}r_{1}r_{2}\cdots r_{n-1}\cdots\cdots r_{1}r_{2}r_{1}$. Denote by $\nc [X]$
the
skew polynomial  ring on generators $X_{ij}$ labelled by all pairs $i,j$
satisfying $1\leq j\leq
n,\;1\leq i\leq n-j+1 $ and defining relations
\[X_{ij}X_{rs}=p^{ij}_{rs}X_{rs}X_{ij},\]
where
\[p^{ij}_{rs}=\left\{\begin{array}{rrr}
q^{2}&\mbox{if}& i>r,\;j=s\\
q&\mbox{if}& i\leq r,\;j=s-1\\
q^{-1}&\mbox{if}& i> r,\;j=s-1\\
1&\mbox{if}& j<s-1 .\end{array}\right.\]

In this  case the map $\Phi=\Phi (w_{0})$ ( here $w_{0}$
stands for the reduced decomposition \newline $w_{0}=r_{1}r_{2}\cdots
r_{n}r_{1}r_{2}\cdots r_{n-1}\cdots\cdots r_{1}r_{2}r_{1}$ ) acts as follows
\begin{equation}
\label{feig_emb_sln}
\Phi(w_{0})(F_{i})= X_{1i}+X_{2i}+\cdots X_{n+1-i\,i}
\;1\leq i\leq n.
\end{equation}

One solves (\ref{feig_emb_sln}) as a system of equations on $X_{ij},\;1\leq
i<j\leq n$ with
coefficients in
$Q(\Phi(U^{-}_{q}(\gtsl_{n+1})))$.

\begin {lemma} The following formulas hold
\[X_{1\,n-1}=\frac{q}{q-q^{-1}}[\Phi(F_{n-1}),\Phi
(F_{n})]_{q}\Phi(F_{n})^{-1},\]
\[X_{2\,n-1}=\frac{1}{q-q^{-1}}[\Phi(F_{n}),\Phi
(F_{n-1})]_{q}\Phi(F_{n})^{-1},\]
\[X_{1i}=\frac{q}{q-q^{-1}}[\Phi(F_{i}),X_{1i+1}]_{q}X_{1i+1}^{-1},\;1\leq i
\leq n-2,\]
%% FOLLOWING LINE CANNOT BE BROKEN BEFORE 80 CHAR
%% FOLLOWING LINE CANNOT BE BROKEN BEFORE 80 CHAR
\[X_{n-i+1\,i}=\frac{1}{q-q^{-1}}[X_{n-i\,i+1},\Phi(F_{i})]_{q}X_{n-i\,i+1}^{-1},\;1\leq i \leq n-2,\]
\[X_{ji}=\frac{1}{q-q^{-1}}(X_{j-1\,i+1}\Phi(F_{i})X_{j-1\,i+1}^{-1}-
X_{j\,i+1}\Phi(F_{i})X_{j\,i+1}^{-1}),
\; 2\leq j\leq n-i .\]
\label{solve_sln}
\end{lemma}
One uses this lemma to prove that the Conjecture~\ref{conj_of_feig} is true.
\begin{theorem}
\label{proof_of_feig}
\mbox{ (i) } The map $\Phi$ is an embedding.

\mbox{ (ii) } The embedding $\Phi: U^{-}_{q}(\gtsl_{n+1})\rightarrow \nc [X]$
induces an isomorphism
\newline $Q(U^{-}_{q}(\gtsl_{n+1}))\approx Q( \nc [X])$.
\end{theorem}

{\em Proof.} Lemma ~\ref{solve_sln} actually shows that
$Q(\Phi(U^{-}_{q}(\gtsl_{n+1})))\approx
Q(\nc[X])$. It is, therefore, enough to prove that $\Phi$ is injective. In
{}~\cite{gk1} Gel'fand and
Kirillov associated a number to an arbitrary algebra $\ca$ which is now known
as the {\em
Gel'fand-Kirillov dimension} $dim_{G-K}\ca$. For example, Gel'fand Kirillov
dimension of a polynomial
ring on $n$ variables, as well as that of the corresponding ring of quotients,
is  equal to $n$. One
of results of ~\cite{gk1} is that if an algebra $\ca$ has a filtration such
that the associated graded algebra
$Gr\ca$ is isomorphic with a polynomial ring on $n$ variables then
$dim_{G-K}\ca=dim_{G-K}Q(\ca)=n$.
Regarding $q$ as an indeterminate and introducing filtration by powers of $q-1$
one derives from
the mentioned results of ~\cite{gk1} their ``$q-$analogues'': dimension
of a ring of skew polynomials on
$n$ indeterminates is equal to $n$ and if $Gr\ca$ is isomorphic with a ring of
skew polynomials on
$n$ indeterminates then $dim_{G-K}\ca=n$. It follows from Proposition
{}~\ref{filter_prop} that
%% FOLLOWING LINE CANNOT BE BROKEN BEFORE 80 CHAR
%% FOLLOWING LINE CANNOT BE BROKEN BEFORE 80 CHAR
$dim_{G-K}U_{q}^{-}(\gtsl_{n+1})=dim_{G-K}\Phi(U_{q}^{-}(\gtsl_{n+1}))=n(n+1)/2$ and, therefore,
$\Phi$ is injective.$\Box$

\subsection {{\bf Complex powers, automorphisms and singular vectors}}

\subsubsection {{\em Construction of automorphisms of quotient rings of
(deformed) universal enveloping algebras and
algebras of skew polynomials} }
\label{constr_of_aut}

{\bf 1.}
For any $ k\in\nn $ the map
 \[\cc_{i}^{k}:\; Q(U_{q}^{-}(\gtg))\rightarrow Q(U_{q}^{-}(\gtg)),\; x\mapsto
F_{i}^{k}xF_{i}^{-k}\]
is an automorphism. Clearly,
$\cc_{i}^{k_{1}+k_{2}}=\cc_{i}^{k_{1}}\circ\cc_{i}^{k_{2}}$. Formulas
(\ref{move_right},
\ref{q_4_mod}) imply that $\cc_{i}^{k}(x)$ is a polynomial function of $k$. For
example, in the
$\gtsl_{n+1}-$case one has
\begin{equation}
\label{simpl_aut}
\cc_{i+1}^{k}(F_{i})=F_{i+1}^{k}F_{i}F_{i+1}^{-k}=
\{k\}F_{i+1}F_{i}F_{i+1}^{-1} +
\{1-k\}F_{i},
\end{equation}
where we have used ``symmetric'' $q-$numbers:
$\{k\}=\frac{q^{k}-q^{-k}}{q-q^{-1}}$.
Using this we define
an automorphism $\cc_{i}^{k}$ by analytic continuation.
 Therefore, with every word $F_{i_{l}}^{\beta_{l}}\cdots F_{i_{1}}^{\beta_{1}}$
we have associated
an automorphism $\cc_{i_{l}}^{\beta_{l}}\cdots \cc_{i_{1}}^{\beta_{1}}$ of
$Q(U_{q}^{-}(\gtg))$.

 {\bf 2.} Since in the case $\gtg=\gtsl_{n+1}$ the rings $Q(U_{q}^{-}(\gtg))$
and $Q(\nc[X])$
are isomorphic with each other
 (  Theorem\ref{proof_of_feig} )
 the above provides the family of automorphisms --
also denoted by
$\cc_{i_{l}}^{\beta_{l}}\cdots \cc_{i_{1}}^{\beta_{1}}$  -- of $Q( \nc [X])$.
  Moreover, the last assertion is valid for any $\gtg$ regardless of
Conjecture~\ref{conj_of_feig}. In reality, there is a construction of
automorphisms of a ring
of skew polynomials which has nothing to do with quantum groups.

Consider for simplicity the ring $\nc[x]=
\nc[x_{1},\ldots,x_{m}]$, $x_{j}x_{i}=q^{2}x_{i}x_{j}$ if $j> i$.
To proceed we need a
{\em $q-$commutative version of the $q-$binomial theorem}:

\begin{equation}
\label{q_comm_vers_mult_th}
(x_{1}+\cdots + x_{m})^{n}=\sum_{i_{1}+\cdots+i_{m}=n}
\frac{[n]!}{[i_{1}]!\cdots [i_{m}]!}x_{1}^{i_{1}}x_{2}^{i_{2}}\cdots
x_{m}^{i_{m}}\;
n\in\nn
 \end{equation}

For $\beta\in\nc$ we set

\begin{eqnarray}
(x_{1}+\cdots + x_{m})^{\beta}=\nonumber\\
\sum_{j=0}^{\infty}\sum_{j_{1}+\cdots + j_{m-1}=j}
\frac{[\beta][\beta-1]\cdots
[\beta-j+1]}{[j_{1}]!\cdots[j_{m-1}]!}x_{1}^{j_{1}}\cdots
x_{r-1}^{j_{r-1}}x_{r}^{\beta
-j} x_{r+1}^{j_{r}}\cdots x_{m}^{j_{m-1}},\label{q_comm_vers_mult_th_compl}
\end{eqnarray}
for some $1\leq r\leq m$, thus making sense out of $(x_{1}+\cdots +
x_{m})^{\beta}$ as an element
of a certain completion of $\nc[x]$ consisting basically of formal power series
(there are exactly $m$ different ways to do that).

Obviously the map $p\mapsto (x_{1}+\cdots + x_{m})^{\beta}p(x_{1}+\cdots +
x_{m})^{-\beta}$ is
an automorphism of the above-mentioned completion. An explicit
calculation ( see below ) shows that
\begin{equation}
\label{aut_of_qcx}
Q(\nc[x])\ni p\;\Longrightarrow (x_{1}+\cdots + x_{m})^{\beta}p(x_{1}+\cdots +
x_{m})^{-\beta}\in
Q(\nc[x])
\end{equation}

Note that the same is true for $(x_{1}+\cdots + x_{m})$ replaced with
$(x_{i_{1}}+\cdots +
x_{i_{k}}),\;1\leq i_{1}<\cdots < i_{k}$ and -- with minor restrictions -- for
$\nc[x]$
replaced with an arbitrary
 ring of skew polynomials. In particular, in the case of the ring $\nc[X]$
related to
$U_{q}(\gtg)$ by the Feigin's construction one obtains automorphisms
\[\cc_{i}^{\beta}p=(\Phi(F_{i}))^{\beta}p(\Phi(F_{i}))^{-\beta}.\]

{\em Remarks.}

(i). It is natural to set $\log F_{i}=\frac{d}{dk}\mid_{k=0}\cc_{i}^{k}$. By
definition $\log F_{i}$
is a differentiation of  $Q(U_{q}^{-}(\gtg))$ as well as of $Q(\nc[X])$. It is
easy to see that, moreover, this is an exterior differentiation. {\em Problem}:
classify non-trivial
 ( exterior modulo inner ) automorphisms of $Q(U_{q}^{-}(\gtg)),\;Q(\nc[X])$.

(ii). The set of  words $F_{i_{l}}^{\beta_{l}}\cdots
F_{i_{1}}^{\beta_{1}},\;\beta_{1},\ldots\beta_{l}\in\nc$  is naturally equipped
with
a group structure. With each such a word one may associate an infinite series:
its expansion over
a ``Poincare-Birkhof-Witt type  basis'' $F^{k}$, where complex powers of
$F_{i}$ are allowed.
(For details in classical setting see ~\cite{malff}.)
Thus we have identified this group with a subgroup of a certain
infinite-dimensional
group with a non-trivial topology. In  ~\cite{kh-z}
 similar group was considered in the classical
case of differential operators on the line. In particular, it was shown that
this is a Poisson-Lie
group.

 {\bf 3.} {\em Calculation of
$(x_{1}+x_{2})^{\beta}x_{2}(x_{1}+x_{2})^{-\beta}$.}
It is easy to see that the proof of (\ref{aut_of_qcx}) reduces to the case
$m=2,\; p=x_{2}$.
One has
\[(x_{1}+x_{2})^{\beta}x_{2}(x_{1}+x_{2})^{-\beta}=
x_{2}(q^{-2}x_{1}+x_{2})^{\beta}(x_{1}+x_{2})^{-\beta}.\]

The $q-$commutative version of the binomial theorem gives
\begin{eqnarray}
%% FOLLOWING LINE CANNOT BE BROKEN BEFORE 80 CHAR
%% FOLLOWING LINE CANNOT BE BROKEN BEFORE 80 CHAR
(q^{-2}x_{1}+x_{2})^{\beta}=q^{-2\beta}x_{1}^{\beta}\sum_{i=0}^{\infty}\frac{(q^{-2\beta})_{i}}
{(q^{2})_{i}}(-q^{2(\beta +1)}x_{1}^{-1}x_{2})^{i} \label{binom_exp_aux_1}\\
(x_{1}+x_{2})^{-\beta}=\{\sum_{i=0}^{\infty}\frac{(q^{2\beta})_{i}}
{(q^{2})_{i}}(-x_{1}^{-1}x_{2})^{i}\}x_{1}^{-\beta}, \label{binom_exp_aux_2}
\end{eqnarray}
where as usual $(a)_{i}=(1-a)(1-aq^{2})\cdots (1-aq^{2(i-1)})$.

In order to show that in the product of the left hand sides of
(\ref{binom_exp_aux_1}-
\ref{binom_exp_aux_2}) almost everything cancels out we employ a {\em
commutative version of
the $q-$binomial theorem} ~\cite{gasp} which reads as follows:
\begin{equation}
\label{comm_vers_qbin_th}
%% FOLLOWING LINE CANNOT BE BROKEN BEFORE 80 CHAR
%% FOLLOWING LINE CANNOT BE BROKEN BEFORE 80 CHAR
\sum_{i=0}^{\infty}\frac{(a)_{i}}{(q^{2})_{i}}z^{i}=\frac{(az)_{\infty}}{(z)_{\infty}},\;z\in\nc,
\end{equation}
where $(a)_{\infty}=\prod_{i\geq 0}(1-aq^{2i})$.

( Although we are in the non-commutative realm the usage of
(\ref{comm_vers_qbin_th}) makes sense
for the right hand sides of (\ref{binom_exp_aux_1}-
\ref{binom_exp_aux_2}) basically involve only one ``variable''
$x_{1}^{-1}x_{2}$.)

By (\ref{comm_vers_qbin_th}) the equalities (\ref{binom_exp_aux_1}-
\ref{binom_exp_aux_2}) are rewritten as follows:

\begin{eqnarray}
(q^{-2}x_{1}+x_{2})^{\beta}&=&q^{-2\beta}x_{1}^{\beta}
\frac{(-q^{2}x_{1}^{-1}x_{2})_{\infty}}{(-q^{2(\beta
+1)}x_{1}^{-1}x_{2})_{\infty}}
 \label{binom_exp_aux_3}\\
(x_{1}+x_{2})^{-\beta}&=&\frac{(-q^{2\beta}x_{1}^{-1}x_{2})_{\infty}}
{(x_{1}^{-1}x_{2})_{\infty}}x_{1}^{-\beta}
 \label{binom_exp_aux_4}
\end{eqnarray}

Carrying out the multiplication one observes that almost all factors of
infinite products cancel
out:

\begin{eqnarray}
(x_{1}+x_{2})^{\beta}x_{2}(x_{1}+x_{2})^{-\beta}=
%% FOLLOWING LINE CANNOT BE BROKEN BEFORE 80 CHAR
%% FOLLOWING LINE CANNOT BE BROKEN BEFORE 80 CHAR
q^{-2\beta}x_{2}x_{1}^{\beta}(1+q^{2\beta}x_{1}^{-1}x_{2})(1+x_{1}^{-1}x_{2})^{-1}x_{1}^{-\beta}=
\nonumber\\
%% FOLLOWING LINE CANNOT BE BROKEN BEFORE 80 CHAR
%% FOLLOWING LINE CANNOT BE BROKEN BEFORE 80 CHAR
q^{-2\beta}x_{2}(1+x_{1}^{-1}x_{2})(1+q^{-2\beta}x_{1}^{-1}x_{2})^{-1}\label{aut_final},
\end{eqnarray}
which completes the proof.

\subsubsection {{\em Application to singular vectors in Verma modules}}
\label{Application_to_singular_vectors_Verma_modules}
It follows from sect.~\ref{constr_of_aut} that elements of the form
\[F_{i_{l}}^{s_{l}}\cdots F_{i_{1}}^{s_{1}}F_{i_{0}}^{N}F_{i_{1}}^{t_{1}}\cdots
F_{i_{l}}^{t_{l}}\]
belong to $Q(U_{q}^{-}(\gtg))$
if $N\in\nz,\;s_{i}+t_{i}\in\nz,\,1\leq i\leq l$. It was shown in
{}~\cite{malff,mal} that such
expressions are relevant to singular vectors in Verma modules. Here we
explicitly calculate them in
the case $\gtg=\gtsl_{n+1}$.

Recall that a
  Verma module $M(\lambda),\; \lambda = (\lambda_{1},\ldots
,\lambda_{n})\in\nc^{n}$ is said to
be a $U_{q}(\gtg)-$module on one generator $v_{\lambda}$ and the following
defining relations

%% FOLLOWING LINE CANNOT BE BROKEN BEFORE 80 CHAR
%% FOLLOWING LINE CANNOT BE BROKEN BEFORE 80 CHAR
\[U^{+}_{q}(\gtg)v_{\lambda}=0,\,K_{i}v_{\lambda}=q_{i}^{\lambda_{i}}v_{\lambda}\,i=1,\ldots,n.\]

 It is easy to see that $M(\lambda)$ is reducible if and only if it contains a
singular vector,
i.e. a non-zero vector different from $v_{\lambda}$ and annihilated by
$U^{+}_{q}(\gtg)$. The
reducibility criterion is the same as in the classical case
{}~\cite{kac_kazhd,dc_kac} and for
 $U_{q}(\gtsl_{n+1})$ reads
as follows:

{\em $M(\lambda)$ is reducible if and only if for some $1\leq i < j\leq
n,\;N\in\nn$
\begin{equation}
\label {k_k_cond}
\lambda_{i}+\lambda_{i+1}+\cdots \lambda_{j}+j-i+1=N.
\end{equation} }

It is known that for a generic point $\lambda$
 on the hyperplane determined by (\ref{k_k_cond}) there is a unique (up to
proportionality)
singular vector in $M(\lambda)$. This means that there is a function sending a
point $\lambda$
on the hyperplane to $S_{ij}^{N}(\lambda)\in U_{q}^{-}(\gtsl_{n+1})$ so
 that the vector $S_{ij}^{N}(\lambda)v_{\lambda}$ is singular. We are going to
evaluate
$S_{ij}^{N}(\lambda)$.

(\ref{k_k_cond}) can be rewritten in the following parametric form
\[\lambda_{j}=N-t_{j}-1,\]
\[\lambda_{j-1}=t_{j}-t_{j-1}-1,\]
\[\lambda_{j-2}=t_{j-1}-t_{j-2}-1,\]
\[\ldots\]
\[\lambda_{i+1}=t_{i+2}-t_{i+1}-1,\]
\[\lambda_{i}=t_{i+1}-1.\]
It follows from ~\cite{mal} that
\begin{equation}
\label{sing_v_form}
S_{ij}^{N}(t) = F_{j}^{t_{j}}\cdots
F_{i+1}^{t_{i+1}}F_{i}^{N}F_{i+1}^{N-t_{i+1}}\cdots
F_{j}^{N-t_{j}}.
\end{equation}
Though (\ref{sing_v_form}) is not quite explicit it is sometimes most
convenient for derivation of
properties of singular vectors. For example, playing with complex powers one
proves that singular
vectors related to $N>1$ are expressed in terms of singular vectors related to
$N=1$. One has
\begin{eqnarray}
S_{ii+1}^{N}(t)=F_{i+1}^{t}F_{i}^{N}F_{i+1}^{N-t}=\nonumber\\
F_{i+1}^{t}F_{i}F_{i+1}^{1-t}F_{i+1}^{t-1}F_{i}F_{i+1}^{2-t}
\cdots F_{i+1}^{t-N+1}F_{i}F_{i+1}^{N-t}=\nonumber\\
S_{ii+1}^{1}(t)S_{ii+1}^{1}(t-1)\cdots S_{ii+1}^{1}(t-N+1).
\end{eqnarray}
Arguing by induction one proves that likewise
\begin{equation}
\label{dep_N}
S_{ij}^{N}(t)=
S_{ij}^{1}(t)S_{ij}^{1}(t-1)\cdots S_{ij}^{1}(t-N+1).
\end{equation}
Therfore, it is enough to calculate $S_{ij}^{1}(t)$. It follows from
(\ref{simpl_aut}) that
\begin{equation}
\label{expl_form_simpl}
S_{ii+1}^{1}(t)=\{t\}F_{i+1}F_{i} +
\{1-t\}F_{i}F_{i+1}.
\end{equation}

Using (\ref{expl_form_simpl}) several times one reduces (\ref{sing_v_form}) to
a form containing
only natural powers of generators.
Denote by $\cp$ the set of all sequences $\epsilon=(\epsilon_{i+1},\ldots
,\epsilon_{j})$,
where each $\epsilon_{m}$ is either 0 or 1. For each $\epsilon\in\cp$ fix a
bijection
$k_{\epsilon}:\{i+1,\ldots ,j\}\rightarrow \{i+1,\ldots ,j\}$ satisfying
\begin{eqnarray}
k^{-1}_{\epsilon}(m)<
k^{-1}_{\epsilon}(m-1)&\mbox{if}&\epsilon_{m}=1,\nonumber\\
k^{-1}_{\epsilon}(m)> k^{-1}_{\epsilon}(m-1)&\mbox{if}&\epsilon_{m}=0.\nonumber
\end{eqnarray}
( Such a bijection obviously exists, though is not unique. However, the final
result is independent
of a choice.)
Further, with each $\epsilon\in\cp$ associate a number $A_{\epsilon}$, given by
\[A_{\epsilon}=\prod_{m=1}^{j-i}\{t_{m,\epsilon}\},\]
where
\[t_{m,\epsilon}=\left\{\begin{array}{ccc}
t_{m}&\mbox{if}&\epsilon_{m+i}=1\\
1-t_{m}&\mbox{if}&\epsilon_{m+i}=0\end{array}\right.\]

\begin{theorem}
%% FOLLOWING LINE CANNOT BE BROKEN BEFORE 80 CHAR
%% FOLLOWING LINE CANNOT BE BROKEN BEFORE 80 CHAR
\[S_{ij}^{1}(t)=\sum_{\epsilon\in\cp}A_{\epsilon}F_{k_{\epsilon}(i)}F_{k_{\epsilon}(i+1)}
\cdots F_{k_{\epsilon}(j)},\]
\[S_{ij}^{N}(t)=
S_{ij}^{1}(t)S_{ii+1}^{1}(t-1)\cdots S_{ij}^{1}(t-N+1).\]
\label{sing_vect_form_expl}
\end{theorem}

\section {{\bf $U_{q}(\gtg)-$modules and $q-$connections}}

\subsection {{\em Modules $U_{q}^{-}[S^{-1}]v_{\lambda}$}}
\label{Modules_U_loc_lambda}
Let $S\subset U_{q}^{-}$ consist of homogeneous elements, and such that
$U_{q}[S^{-1}]$ is
well-defined. A typical example of $S$ is a multiplicative span of an arbitrary
subset of
$\{F_{1},\ldots,F_{n}\}$.
 The following isomorphism of vector spaces is an analogue of the triangular
decomposition:
\[U_{q}[S^{-1}]\approx U_{q}^{-}[S^{-1}]\otimes U_{q}^{\geq} .\]
(Existence of this isomorphism follows from the relation \newline $[E_{i},S_{j}
^{-1}]=-S_{j}^{-1}[E_{i},S_{j}]S_{j}^{-1}$, which allows
to commute $E'$s to the right.) Denote by $\nc_{\lambda}$ a character of $
U_{q}^{\geq}$ defined by $E_{i}\rightarrow 0,\; K_{i}\rightarrow
q^{\lambda_{i}};\;1\leq i\leq n$.
A Verma module over $U_{q}[S^{-1}]$ is said to be
$U_{q}[S^{-1}]\otimes_{U_{q}^{\geq}}\nc_{\lambda}$.
Denote by $v_{\lambda}$ the image of $1\otimes 1$ in
$U_{q}[S^{-1}]\otimes_{U_{q}^{\geq}}\nc_{\lambda}$.  Clearly,
$U_{q}[S^{-1}]\otimes_{U_{q}^{\geq}}\nc_{\lambda}$ is a free
$U_{q}^{-}[S^{-1}]-$module generated
by $v_{\lambda}$. We shall be interested in the restriction of
$U_{q}[S^{-1}]\otimes_{U_{q}^{\geq}}\nc_{\lambda}$ to $U_{q}(\gtg)$. Due to the
lack of better
notation $U_{q}^{-}[S^{-1}]v_{\lambda}$ will stand for this restriction.

Note that the module $U_{q}^{-}[S^{-1}]v_{\lambda}$ is always reducible for it
contains a Verma
 module $M(\lambda)=U_{q}^{-}v_{\lambda}$. Though its structure is unknown in
general, we are
able to describe it in the simplest case when $S$ is multiplicatively generated
by one of $F'$s, say,
$F_{i}$.

It is easy to see that
\begin{equation}
\label{actofE}
%% FOLLOWING LINE CANNOT BE BROKEN BEFORE 80 CHAR
%% FOLLOWING LINE CANNOT BE BROKEN BEFORE 80 CHAR
E_{j}F_{i}^{m}v_{\lambda}=\delta_{ij}\{m\}F_{i}^{m-1}\frac{q_{i}^{\lambda_{i}-m+1}-q_{i}^{-\lambda_{i}+m-1}}
{q_{i}-q_{i}^{-1}}v_{\lambda},\;m\in\nz.\end{equation}

One realizes that
$U_{q}^{-}[F_{i}^{-1}]v_{\lambda}$ is a module induced from the representaion
of the
parabolic subalgebra generated by $E_{1},\ldots,E_{n},F_{i}$
in the space $\oplus_{m\in\nz}F_{i}^{m}v_{\lambda}$. Further, if $\lambda_{i}$
is not in
$\{-2,-3,\ldots\}$ then (\ref{actofE}) implies that $E_{i}$ acts freely on the
quotient module
$U_{q}^{-}[F_{i}^{-1}]v_{\lambda}/M(\lambda)$. It is now easy to show that
$U_{q}^{-}[F_{i}^{-1}]v_{\lambda}/M(\lambda)$ is a Verma module related to a
 Borel subalgebra  $R_{i}U_{q}^{\geq}$ twisted by the Lusztig's automorphism
 ( see
(\ref{luszt_aut_1},\ref{luszt_aut_2},\ref{luszt_aut_3}) )
 and the highest weight $\lambda+\alpha_{i}$. If, however, $\lambda_{i}$ does
belong to
$\{-2,-3,\ldots\}$ then, as (\ref{actofE}) implies, the vector
$F_{i}^{\lambda_{i}+1}$ is singular.
This means that there arises a chain of submodules $M(\lambda)\subset
M(\lambda+(\lambda_{i}+1)\alpha_{i})\subset U_{q}^{-}[F_{i}^{-1}]v_{\lambda}$.
As above one shows
that the quotient module
$U_{q}^{-}[F_{i}^{-1}]v_{\lambda}/M(\lambda+(\lambda_{i}+1)\alpha_{i})$
is a Verma module related to a twisted Borel subalgebra and the highest weight
$\lambda+(\lambda_{i}+2)\alpha_{i})$.

For the sake of breavity, denote by $^{R_{i}}M_{q}(\lambda)$ an
$U_{q}(\gtg)-$module, isomorphic to
$M(\lambda)$ as a vector space with the action being twisted by $R_{i}$:

\[U_{q}(\gtg)\ni x\mapsto R_{i}x\mapsto End(M(\lambda)).\]
 We have obtained
\begin{proposition}
\label{str_F-1}
If $\lambda_{i}$ is not in
$\{-2,-3,\ldots\}$ then\newline
$U_{q}^{-}[F_{i}^{-1}]v_{\lambda}/M(\lambda)\approx\, ^{R_{i}}
M(\lambda+\alpha_{i})$.

If $\lambda_{i}\in\{-2,-3,\ldots\}$ then there exists a chain of submodules
$M(\lambda)\subset
M(\lambda+(\lambda_{i}+1)\alpha_{i})\subset U_{q}^{-}[F_{i}^{-1}]v_{\lambda}$
and
\newline
$U_{q}^{-}[F_{i}^{-1}]v_{\lambda}/M(\lambda+(\lambda_{i}+1)\alpha_{i})
\approx\, ^{R_{i}} M(\lambda+(\lambda_{i}+2)\alpha_{i})$.
\end{proposition}

Observe that Proposition ~\ref{str_F-1} along with its proof carries over to
the case of a quantum
group attached to an arbitrary symmetrizable Cartan matrix $A$.

\subsection{ {\em Modules realized in skew polynomials}}
{\bf 1.}The Feigin's embedding $U_{q}^{-}\rightarrow \nc^{s}[X]$ makes the
latter into a
$U_{q}^{-}-$module, action being defined by means of the left multiplication.
One may want to extend
this to an action of the entire $U_{q}$. It is straightforward in view of the
results of the
previous section if $\gtg=\gtsl_{n+1}$ for in this case $Q(\nc[X])\approx
Q(U_{q})$
(Theorem~\ref{proof_of_feig}) and one obtains a family of modules
$Q(\nc[X])v_{\lambda}\;(=Q(U_{q})v_{\lambda})$. The module
$Q(\nc[X])v_{\lambda}$ is definitely too
big and it is natural to confine to the smallest submodule containing
$\nc[X]v_{\lambda}$. This
module is still always reducible, for example, it contains a Verma module
$M_{q}(\lambda)$ -- the one
generated by $X_{1i}+\cdots + X_{n-i+1,i},\;1\leq i\leq n$ -- and
$U_{q}^{-}[F_{n}^{-1}]v_{\lambda}$
-- the one generated by $X_{1i}+\cdots + X_{n-i+1,i},\;1\leq i\leq
n-1,\;X_{1n}^{\pm 1}$
 Though we do not have an explicit description of this module in
general we are able to consider the case of $\gtsl_{3}$ in full detail.

\begin{proposition}
\label{feig_mod_sl3}
For generic $\lambda$
$U_{q}(\gtsl_{3})\cdot\nc[X_{11},X_{21},X_{12}]v_{\lambda}=
\nc[X_{11},X_{21},X_{12}^{\pm 1}]v_{\lambda}$. For any $\lambda$
$\nc[X_{11},X_{21},X_{12}^{\pm 1}]v_{\lambda}\approx
U_{q}^{-}(\gtsl_{3})[F_{2}^{-1}]v_{\lambda}$.
\end{proposition}

Proposition~\ref{str_F-1}, therefore, determines the structure of
 $\nc[X_{11},X_{21},X_{12}^{\pm}]v_{\lambda}$.

As to the general case, the module in question should also be isomorphic to a
module
$U_{q}^{-}[S^{-1}]v_{\lambda}$ for an appropriate set $S$ determined by
formulas of
 Lemma~\ref{solve_sln}.

{\bf 2.} The above is relevant to the Gel'fand-Kirillov conjecture for
$U_{q}(\gtsl_{n+1} )$.
Denote by $\cd[X]$ an algebra of q-difference operators acting on $Q(\nc[X])$.
In other words, $\cd[X]$ is an algebra generated by $Q(\nc[X])$
viewed as operators
of left multiplication and $T_{ij},\;1\leq j\leq n,\,1\leq i\leq n-j+1$, where

\[T_{ij}:\;X_{rs}\mapsto\left\{\begin{array} {lll}
qX_{rs}&\mbox{if}&(i,j)=(rs)\\
X_{rs}&\mbox{if}&(i,j)\neq(rs).\end{array}\right.\]
 Denote by $\cd[X,\lambda]$ the trivial central extension of $\cd[X]$ by
commuting variables
$q^{\lambda_{1}},\ldots , q^{\lambda_{n}}$, where
$\lambda=(\lambda_{1},\ldots,\lambda_{n})$ is
understood as a highest weight. The definition of the module
$Q(\nc[X])v_{\lambda}$ implies that
there exists a family of embeddings -- parametrized by $\lambda$ -- of
$U_{q}(\gtsl_{n+1})$ into
$\cd[X]$ or, equivalently, an embedding
\begin{equation}
\label{q_group_embed}
\rho:\;U_{q}(\gtsl_{n+1})\rightarrow \cd[X,\lambda].
\end{equation}

\begin{conjecture} $\rho$ provides an isomorphism of a quotient field a certain
algebraic
extension $\hat{U}_{q}(\gtsl_{n+1})$
 of $U_{q}(\gtsl_{n+1})$ with a quotient field of a certain subalgebra of
$\cd[X,\lambda]$.
\end{conjecture}

 Construction of $\hat{U}_{q}(\gtsl_{n+1})$, which goes back to Gelfand and
Kirillov ~\cite{gk2}, is as follows:

Identify $U_{q}(\gtsl_{n+1})$ with
$\rho(U_{q}(\gtsl_{n+1}))\subset\cd[X,\lambda]$ and define
$\hat{U}_{q}(\gtsl_{n+1})$ to be the subalgebra of $\cd[X,\lambda]$ generated
by
$U_{q}(\gtsl_{n+1})$, $q^{\lambda_{1}},\ldots , q^{\lambda_{n}} $ and
$K^{\omega_{1}},\ldots,
K^{\omega_{n}}$, where $\omega_{i},\;1\leq i\leq n$ are dual fundamental
weights, i.e.
$\alpha_{j}(\omega_{i})=\delta_{ji}$. ( It is meant that
$K^{\omega}E_{i}=q_{i}^{\alpha_{i}(\omega)}E_{i}K^{\omega}$.) Note that
elements
$q^{\lambda_{1}},\ldots , q^{\lambda_{n}} $ generate a certain finite algebraic
extension of the
center of $U_{q}(\gtsl_{n+1})$. ( Description of the center of
$U_{q}(\gtsl_{n+1})$ may be found in
{}~\cite{dc_kac}.)

We have been able to verify the conjecture in the cases of $\gtsl_{2}$,
$\gtsl_{3}$ by
straightforward calculation of $\rho^{-1}$,
 which is simple in the $\gtsl_{2}$-case and rather tiresome in the
$\gtsl_{3}$-case.

\begin{proposition}

\[\mbox{ (i) } Q(\hat{U}_{q}(\gtsl_{2}))\approx \cd[X,\lambda].\]
(In this case $X$ stands for $X_{11}$.)

\mbox{ (ii) } $Q(\hat{U}_{q}(\gtsl_{3}))$ is isomorphic with the quotient field
of the subalgebra
 of $\cd[X,\lambda]$ generated by $T_{11}^{2},T_{21}^{2},T_{11}T_{21}
, T_{12};\;X_{11},X_{21},X_{12};\;q^{\lambda_{1}},q^{\lambda_{2}}$.
\end{proposition}

\subsection {{\bf $U_{q}(\gtg)-$modules and $q-$connections.}}
\subsubsection{{\em A $q-$connection with coefficients in a ring of skew
polynomials.}}
Let $\nc[x]:=\nc[x_{1},\ldots,x_{n}],\;x_{j}x_{i}=q^{2}x_{i}x_{j},\;i<j$ be a
ring of skew polynomials
(as yet it has nothing to do with quantum groups) and  $\cd[x]$ be
the corresponding ring of $q-$difference
operators.

By a {\em quantum line bundle} we mean a free rank 1 module over $\nc[x]$ or,
more generally,
$\nc[x][S^{-1}$ for a suitable $S$. Sections of a quantum line bundle, i.e.
elements of
$\nc[x][S^{-1}]$, therefore become a $\cd[x]$-module. In other words there
arises an inclusion
\[\nabla^{triv}:\; \cd[x]\hookrightarrow\,End(\nc[x][S^{-1}]).\]

By a {\em $q-$connection} with
coefficients in a quantum line bundle we mean an associative algebra
homomorphism $\nabla:\;
\cd[x]\rightarrow\,End(\nc[x][S^{-1}])$  such that
$\nabla(x_{i})=L(x_{i}),\;\nabla(T_{i})=R(b_{i}(x))T_{i},\;1\leq i\leq n$,
where $b_{i}(x)\in
Q(\nc[x])$ and $L(x_{i}),\;R(b_{i}(x))$ stand for the operator of the left or
right (resp.)
multiplication by $x_{i}$ or $b_{i}(x)$ (resp.).

The same can be equivalently described in terms of cohomology. For
$\chi=(\chi_{1},\ldots,\chi_{n})\in\nz^{n}$ set
$T^{\chi}=T_{1}^{\chi_{1}}\circ\cdots\circ
T_{n}^{\chi_{n}}$ and $\nabla_{\chi}=\nabla(T^{\chi})$. Obviously,
$\nabla_{\chi}=R(b_{\chi}(x))T^{\chi}$ for some $b_{\chi}(x)\in Q(\nc[x])$. The
associative algebra
homomorphism condition reads as
\[b_{\chi_{1}+\chi_{2}}(x)=(T^{\chi_{1}}b_{\chi_{2}}(x))b_{\chi_{1}}(x).\]
The last equality simply means that the map $\nz^{n}\ni\chi\mapsto
b_{\chi}(x)\in Q(\nc[x])$ is a
1-cocycle of an abelian group $\nz^{n}$ with coefficients in $Q(\nc[x])$. In
this language the above
$\nabla^{triv}$ is related to the cocycle $\chi\mapsto 1$.

 It is natural to say that
a cocycle is trivial if it is given by
$b_{\chi}(x)\,=\,(T^{\chi}r(x))r^{-1}(x)$ for some
$r(x)\in Q(\nc[x])$.Indeed the cocycle $\chi\mapsto (T^{\chi}r(x))r^{-1}(x)$
makes into
the cocycle $\chi\mapsto 1$ by ``the change of trivialization'': $f(x)\mapsto
f(x)r(x)^{-1}$.
 The cocycle $\chi\mapsto (T^{\chi}r(x))r^{-1}(x)$ is a coboundary of the
0-cocycle $r(x)$.
Therefore we have established a 1-1 correspondence between non-trivial
$q-$connections and elements
of $H^{1}(\nz^{n},\;Q(\nc[x]))$.

To produce a construction of some elements of $H^{1}(\nz^{n},\;Q(\nc[x]))$
 fix arbitrary subsets $J_{1},\ldots , J_{l}$ of $\{1,\ldots ,m\}$ and set
\[r_{i}=\sum_{j\in J_{i}}x_{j}.\]
Let $\Psi=r_{1}^{\beta_{1}}r_{2}^{\beta_{2}}\cdots r_{l}^{\beta_{l}}$ for
some $\beta_{1},\ldots,\beta_{l}\in\nc$ ( see sect.\ref{constr_of_aut} ).

\begin{lemma}
\label{constr_of_el_H1}
The correspondence
$\chi\mapsto \Psi_{\chi}=(T^{\chi}\Psi)\Psi^{-1}$ represents an element of
$H^{1}(\nz^{n},\;Q(\nc[X]))$.
\end{lemma}

{\bf Proof.}
The fact that $\Psi_{\chi}=(T^{\chi}\Psi)\Psi^{-1}$ is a 1-cocycle is obvious
for this is a
coboundary of $\Psi$. (One may also think of it as  the ``local'' change of
trivialization
$f(x)\mapsto f(x)\Psi(x)$ in the bundle with the trivial $q-$connection.)
 What has to be proven is that  $\Psi_{\chi}\in Q(\nc[X])$ for any $\chi$.
To do this observe that

\[x_{i}^{-1}\frac{1-T^{2}_{i}}{1-q^{2}}(x_{1}+\cdots + x_{m})^{\beta}=
[\beta](q^{-2}x_{1}+\cdots +q^{-2}x_{i-1}+x_{i}+\cdots + x_{m})^{\beta-1}.\]
The calculation as in
(\ref{binom_exp_aux_3},\ref{binom_exp_aux_4},\ref{aut_final}) shows that
$(q^{-2}x_{1}+\cdots +q^{-2}x_{i-1}+x_{i}+\cdots + x_{m})^{\beta-1}=p
(x_{1}+\cdots + x_{m})^{\beta},\;p\in Q(\nc[x])$.
It implies that
\[T_{i}^{2}(x_{1}+\cdots + x_{m})^{\beta}=\{1-(1-q^{2})[\beta]x_{i}p\}
(x_{1}+\cdots + x_{m})^{\beta}.\]

Therefore, a $q-$ difference operator makes $\Psi$ into
$p_{1}r_{1}^{\beta_{1}}\cdots
p_{l}r_{l}^{\beta_{l}}$
for some $p_{1},\ldots,p_{l}\in Q(\nc[x])$.
To rewrite the latter in the form $p\Psi,\;p\in Q(\nc[x])$
one wants to move each $p_{i}$ to the left. This can be done by using the
automorphisms
\[Q(\nc[x])\ni q\mapsto r_{i}^{\beta}qr_{i}^{-\beta},\]
see (\ref{aut_of_qcx}).
One has
\[p_{1}r_{1}^{\beta_{1}}\cdots
p_{l}r_{l}^{\beta_{l}}=p_{1}\tilde{p_{2}}\cdots\tilde{p_{l}}\Psi,\]
where $\tilde{p_{j}}=
r_{1}^{\beta_{1}}\cdots r_{j-1}^{\beta_{j-1}}p_{j}r_{j-1}^{-\beta_{j-1}}\cdots
p_{1}^{-\beta_{1}},\;2\leq j\leq l$. $\Box$

We will denote by $\nabla (\Psi)$ the connection $\chi\mapsto
\Psi_{\chi}=(T^{\chi}\Psi)\Psi^{-1}$
given by Lemma ~\ref{constr_of_el_H1}.

In the classical case tensor product of a pair of trivial line bundles with
flat connections is
a trivial line bundle equipped with a canonical flat connection. This gives an
operation on
connections.  In classical case  connections are also identified with a certain
1st cohomology
group and this operation happens to be simply an addition.  Though we are
unable to carry out the
same in full generality in the $q-$commutative realm, we can produce the
following  non-commutative
operation on the {q-}connections of the form $\nabla (\Psi)$:
\[\nabla(\Psi_{1})\,
_{q}\!\otimes\nabla(\Psi_{2})=\nabla(\Psi_{1}\Psi_{2}).\]
 This operation is obviously a $q-$analogue
of an addition of 1-cocycles in the classcal setting.

{\em Remark.}
Our approach here is a $q-$ commuative version of that of Aomoto and Kato
 in ~\cite{aomok}.
In particular the construction of cocycles in Lemma ~\ref{constr_of_el_H1} has
its commuative
counterpart, which is claimed to possess a sort of universality property. The
same may be true --
with minor modifications -- in our case.

\subsubsection{{\em $U_{q}(\gtg)$-modules twisted by a $q-$connection.}}
 \label{som_el_of H1_q_case}

 {\bf 1.} {\em Intertwining operators.}
Of course everything written in the previous section applies to more general
rings of skew
polynomials provided one takes more care about the choice of elements
$r_{1},\ldots,r_{l}$.
For example, in the case of an algebra $\nc[X]$, coming from the Feigin's
morphism
$\Phi:\;U_{q}^{-}\rightarrow \nc[X]$ ( see sect.~\ref{gel_kir_feig} ), a
natural choice is
$r_{j}=\Phi(F_{i_{j}})$ for an arbitrary sequence $i_{1},\ldots,i_{l}$. ( There
are some others
which one can easily think of.) In the case of $\gtg=\gtsl_{n+1}$ the
$U_{q}(\gtsl_{n+1})-$module
structure on $Q(\nc[X])$ implies a homomorphism ( see (\ref{q_group_embed}) ):
\begin{equation}
\label{embed_dep_on_lambda}
\rho_{\lambda}:\;U_{q}(\gtsl_{n+1})\rightarrow \cd[X].\end{equation}
 Given a $q-$connection $\nabla$ one twists this
$U_{q}(\gtsl_{n+1})-$module
structure by
\begin{equation}
\label{embed_dep_on_lambda_twist}
\nabla\circ\rho_{\lambda}: \;U_{q}(\gtsl_{n+1})\rightarrow \cd[X].
\end{equation}
As above
denote by $\nabla (\Psi)$ the $q-$connection coming from
\newline$(T^{\chi}\Psi)\Psi^{-1}\in
H^{1}(\nz^{N},\,Q(\nc[X]))$ ,  $\Psi=(\Phi(F_{i_{1}}))^{\beta_{1}}\cdots
(\Phi(F_{i_{l}}))^{\beta_{l}}$ for some $\beta_{1},\ldots,\beta_{l}\in\nc$.
Let $r_{i}\in W$ be a reflection at the simple root $\alpha_{i}$.
Set
\begin{equation}
\label{spec_choice_b}
\beta_{j}=\frac{2(r_{i_{l+2-j}}\cdots
r_{i_{l}}\cdot\lambda,\alpha_{i_{l+1-j}})}
{(\alpha_{i_{l+1-j}},\alpha_{i_{l+1-j}})}+1,
\end{equation}
where $r_{i}\cdot\lambda$ stands for the shifted action of the Weyl group.
Set $\nabla (r_{i_{1}}\cdots r_{i_{l}};\lambda)=\nabla (\Psi)$ if
$\beta_{1},\ldots ,\beta_{l}$ are
given by (\ref{spec_choice_b}).

\begin{proposition}
\label{q_def_of_int_int_op}
There is a $U_{q}(\gtsl_{n+1})-$linear map
of the module related to $\nabla(\Psi)\circ\rho_{w\cdot\lambda}$ into the one
related to
$\nabla(\Psi)\,_{q}\!\otimes\nabla (r_{i_{1}}\cdots
r_{i_{l}};\lambda)\circ\rho_{\lambda}$,
where $w=r_{i_{1}}\cdots r_{i_{l}}$.
\end{proposition}

{\em Proof} Passage from $\rho_{\mu}$ to $\nabla(\Psi)\circ\rho_{\mu}$ means
that one replaces
$v_{\mu}$ (i.e. unit of $\nc[X])$) with $F_{i_{1}}^{\beta_{1}}\cdots
F_{i_{l}}^{\beta_{l}}v_{\mu}$. (See the proof of Lemma ~\ref{constr_of_el_H1}.)
 Formula (\ref{actofE}) implies that under the choice
(\ref{spec_choice_b}) the vector $F_{i_{1}}^{\beta_{1}}\cdots
F_{i_{l}}^{\beta_{l}}v_{\lambda}$ is singular of the weight $
r_{i_{1}}\cdots r_{i_{l}}\cdot\lambda$ and, therefore, satisfies all the
conditions imposed on
$v_{w\cdot\lambda}$. $\Box$

Observe that having looked over the definitions one can make a precise sense
out of the statement:

the module related to $\nabla(\Psi)\circ\rho_{w\cdot\lambda}$ embeds
into the one related to
$\nabla(\Psi)\,_{q}\!\otimes\nabla (r_{i_{1}}\cdots
r_{i_{l}};\lambda)\circ\rho_{\lambda}$ as a space
of sections satisfying a certain regularity condition.

 {\bf 2.} {\em Structure of modules}
$U_{q}^{-}[F_{i}^{-1}]F_{i}^{\beta}v_{\lambda}$. Here we
obtain the structure description of the modules twisted by a $q-$connection in
the simplest case of $\nabla=\nabla((\Phi(F_{i}))^{\beta}),\;\beta\in\nc$. This
module always
contains  a submodule generated by  $1\in\nc[X]$ and, as one easily sees,
isomorphic with the
following extension of a Verma module:
$U_{q}^{-}[F_{i}^{-1}]F_{i}^{\beta}v_{\lambda}$. The
$U_{q}(\gtg)-$module structure on the latter is defined as follows:

(i) $F_{1},\ldots ,F_{n}$ act by left multiplication;

(ii) action of $E_{1},\ldots ,E_{n}$ is determined by setting ( c.f.
(~\ref{actofE}) )
\begin{equation}
\label{actofE_compl}
%% FOLLOWING LINE CANNOT BE BROKEN BEFORE 80 CHAR
%% FOLLOWING LINE CANNOT BE BROKEN BEFORE 80 CHAR
E_{j}F_{i}^{\beta}v_{\lambda}=\delta_{ij}\{\beta\}F_{i}^{\beta-1}\frac{q_{i}^{\lambda_{i}-\beta+1}
-q_{i}^{-\lambda_{i}+\beta-1}}
{q_{i}-q_{i}^{-1}}v_{\lambda},\;\beta\in\nc.
\end{equation}

Denote by $^{i}U_{q}^{+}$ the parabolic subalgebra of $U_{q}(\gtg)$ generated
by \newline
$E_{1},\ldots,E_{n};K_{1}^{\pm 1},\ldots K_{n}^{\pm 1};F_{i}$. The equation
(~\ref{actofE_compl})
determines a structure of $^{i}U_{q}^{+}$-module on the space spanned by
$F_{i}^{\beta+k},\;k\in\nz$.
Denote a  module  obtained in this way by $\cv^{i}_{\lambda}$. Clearly,
$U_{q}^{-}[F_{i}^{-1}]F_{i}^{\beta}v_{\lambda}$ is isomorphic with the unduced
module
$Ind_{^{i}U_{q}^{+}}^{U_{q}}\cv^{i}_{\lambda}$. This isomorphism provides a
precise analogy
 between $U_{q}^{-}[F_{i}^{-1}]F_{i}^{\beta}v_{\lambda}$ and a Verma module
$U_{q}^{-}v_{\lambda}$: one is obtained from another by replacing the vacuum
vector $v_{\lambda}$
with the {\em vacuum chain} $\cv^{i}_{\lambda}$. This analogy can be pushed
further by remarking
that $U_{q}^{-}[F_{i}^{-1}]F_{i}^{\beta}v_{\lambda}$ is reducible if and only
if it contains
a {\em singular chain} in much the same way as a Verma module is reducible if
and only if it
contains a singular vector. Here by a singular chain we naturally mean a
non-zero
$^{i}U_{q}^{+}$-linear map $\cv^{i}_{\mu}\subset
U_{q}^{-}[F_{i}^{-1}]F_{i}^{\beta}v_{\lambda}$
different from $\cv^{i}_{\lambda}\subset
U_{q}^{-}[F_{i}^{-1}]F_{i}^{\beta}v_{\lambda}$. A weight
lattice of a singular chain $\cv^{i}_{\mu}$
 is of the form $\mu+\nz\alpha_{i}$. By the weight of a singular chain
$\cv^{i}_{\mu}$ we mean
an element $\bar{\mu}\in\,\gth^{\ast}/\nz\alpha_{i}$, where $\bar{\mu}$ stands
for an image of $\mu$
under the natural projection $\gth^{\ast}\rightarrow\gth^{\ast}/\nz\alpha_{i}$.
The following
partially relies  on sect.~\ref{Modules_U_loc_lambda}.

\begin{theorem}
\label{str_F-1_tw}
(i)If $\beta\in\nz$ or $\beta\in\lambda(H_{i})+\nz$ then
$U_{q}^{-}[F_{i}^{-1}]F_{i}^{\beta}v_{\lambda}$ is isomorphic with either
$U_{q}^{-}[F_{i}^{-1}]v_{\lambda}$ or
$U_{q}^{-}[F_{i}^{-1}]v_{r_{i}\cdot\lambda}$ (resp.);
see Proposition ~\ref{str_F-1}.

(ii) Otherwise $U_{q}^{-}[F_{i}^{-1}]F_{i}^{\beta}v_{\lambda}$ is reducible
$\Longleftrightarrow$
it contains a singular chain of the weight $\overline{\lambda-N\alpha}$ for
some
$\alpha\in\Delta_{+},\;\alpha\neq\alpha_{i},\;N\in\nn$ $\Longleftrightarrow$
there is $j\in\nz,\;N\in\nn$ such that
%% FOLLOWING LINE CANNOT BE BROKEN BEFORE 80 CHAR
%% FOLLOWING LINE CANNOT BE BROKEN BEFORE 80 CHAR
\[(\lambda+\rho,\alpha+j\alpha_{i})=\frac{N}{2}(\alpha+j\alpha_{i},\alpha+j\alpha_{i}),\;
\alpha+j\alpha_{i}\in\Delta_{+},\]
where $\rho\in\gth^{\ast}$ is determined by $\rho(H_{k})=1,\;1\leq k\leq n$.
 \end{theorem}

{\em Remark.} The Verma module $M(\lambda)$  contains a singular vector of the
weight
$\lambda - N\alpha,\;\alpha\in\Delta_{+}$ if and only if $\lambda$ belongs to
the Kac-Kazhdan
hyperplane (~\cite{kac_kazhd},
 see also sect.~\ref{Application_to_singular_vectors_Verma_modules})
related to the pair $(\alpha, N)$:
\begin{equation}
\label{k_k_red_crit}
(\lambda+\rho,\alpha)=\frac{N}{2}(\alpha,\alpha),\;
\alpha\in\Delta_{+}.
\end{equation}
 Item (ii) of Theorem \ref{str_F-1_tw} claims that
$U_{q}^{-}[F_{i}^{-1}]F_{i}^{\beta}v_{\lambda}$ contains a singular chain of
the weight
$\overline{\lambda-N\alpha}$ if and only if $\lambda$ belongs to the union of
Kac-Kazhdan
hyperplanes related to all pairs $(\alpha+j\alpha_{i},
N),\;\alpha+j\alpha_{i}\in\Delta_{+}$.
Therefore a singular chain encodes information on a collection of singular
vectors in a Verma module.

{\em Proof of Theorem ~\ref
{str_F-1_tw}.} Item (i)  immediately follows from definitions and
(~\ref{actofE_compl}). As to
(ii), fix root vectors $E_{\alpha},\;\alpha\in\Delta_{+}$ as in
(\ref{def_root_vect_1}). It is clear
that the space spanned by all singular chains coincides with the space of
solutions of the following
system of linear equation
\begin{equation}
\label{eq_on_sing_chain}
E_{\alpha}w=0\;\mbox{ for all }\alpha\neq\alpha_{i}
\end{equation}
(This system should be regarded as resticted to each weight space of the
module.)

One deduces from Proposition ~\ref{filter_prop} the space of solutions to
(~\ref{eq_on_sing_chain}) is
a $<F_{i},K_{i}^{\pm1},E_{i}>-$module from which one easily extracts a singular
chain.

All the vector spaces $U_{q}^{-}[F_{i}^{-1}]F_{i}^{\beta}v_{\lambda}$
parametrized by
$\beta\in\nc$ are naturally isomorphic with each other and  with the space
$U_{q}^{-}[F_{i}^{-1}]$. Therefore
(~\ref{eq_on_sing_chain}) may be regarded as a family of  systems of linear
 equations on $U_{q}^{-}[F_{i}^{-1}]$
 polynomially depending on $\beta$. For a fixed
weight space of $U_{q}^{-}[F_{i}^{-1}]$ existence of solutions to
(~\ref{eq_on_sing_chain})
 lying in this space is equivalent to vanishing of a certain polynomial
depending
on $\lambda$ and $\beta$. It is easy to deduce, however, that under the
assumptions of (ii)  once
there is a solution for $\beta=\beta_{0}$ then there are solutions for infinite
many values
$\beta\in\beta_{0}+\nz$. Therefore the mentioned polynomial is actually
independent of $\beta$. Now we
may set $\beta=0$ without lack of generality.
But then it is easy to see that -- again under the assumptions of (ii) --
 the same system (~\ref{eq_on_sing_chain}) gives reducibility criterion
for the submodule $M(\lambda)$ of $U_{q}^{-}[F_{i}^{-1}]v_{\lambda}$, see
Proposition ~\ref{str_F-1}. The proof now follows from the Kac-Kazhdan
equations, see the above
Remark. $\Box$

The Kac-Kazhdan reducibility criterion (~\ref{k_k_red_crit}) was carried over
to the case of a quantum
group attached to an arbitrary symmetrizable Cartan matrix $A$ in ~\cite{mal}.
It follows that
Theorem ~\ref{str_F-1_tw}  remains valid in this general setting.

\end{document}